\documentclass[sigconf,nonacm]{acmart}

\renewcommand\footnotetextcopyrightpermission[1]{}

\usepackage{booktabs}
\usepackage{listings}
\usepackage{xspace}

\usepackage[htt]{hyphenat}

\begin{document}

\title{An Assessment Framework for \\ Application-Level Cryptographic Agility}

\author{Navaneeth Rameshan}
\email{vme@zurich.ibm.com}
\affiliation{\institution{IBM Research Europe}\city{Zurich}\country{Switzerland}}

\author{Gregoire Messmer}
\email{Gregoire.Messmer@ibm.com}
\orcid{0000-0000-0000-0000}
\affiliation{\institution{IBM Research Europe}\city{Zurich}\country{Switzerland}}

\renewcommand{\shortauthors}{Rameshan and Messmer}

\begin{abstract}

The impending post-quantum transition to new cryptography will require complete replacement of algorithms within all software. The cryptographic APIs used today make this transition challenging because they were not designed with agility as a concern. There is no method for systematically assessing \textit{cryptographic agility} as an overall ability. In addition to this, the term itself refers to multiple independent capabilities. Specifically, it includes replacing algorithms, selecting by policy, and substituting implementations. This lack of structured decomposition limits both the evaluation of systems and the development of cryptographically agile APIs.

We introduce a component-based assessment framework that characterizes application-level cryptographic agility along seven orthogonal dimensions: three coupling dimensions that measure what the application code knows about algorithms and providers, a cross-cutting decoupling mechanism, a governance authority dimension, and two agility enablers that measure actual migration capability. The framework is non-linear and captures non-hierarchical profiles: a system may achieve high operation decoupling yet low creation decoupling, or strong versioning without externalized configuration.

We evaluate six representative APIs (PKCS\#11, OpenSSL~3.0, JCA, Google Tink, AWS KMS, and HashiCorp Vault Transit) against the framework, revealing three pervasive and independent gaps: no system supports intent-based key creation, none provides policy-driven algorithm selection (as distinct from access control), and none offers dedicated/first-class operations for algorithm transformation of existing keys. These gaps are individually sufficient to prevent agile migration, explaining why the post-quantum transition remains a software engineering problem despite decades of API progress. 

\end{abstract}


\maketitle

\section{Introduction}
\label{sec:introduction}

The standardization of post-quantum cryptographic algorithms by the National Institute of Standards and Technology (NIST) in 2024 has initiated what may constitute the most extensive cryptographic transition in the history of deployed systems. FIPS~203 (ML-KEM)~\cite{fips203}, FIPS~204 (ML-DSA)~\cite{fips204}, and FIPS~205 (SLH-DSA)~\cite{fips205} establish new cryptographic primitives for key encapsulation and digital signatures, which must eventually replace RSA and elliptic curve cryptography across billions of devices and applications. NIST has announced the deprecation of 112-bit security-strength classical algorithms by 2030 and their disallowance by 2035~\cite{nist-sp800-131a,nist-sp1800-38}, establishing firm timelines that organizations must meet regardless of their software development capacity.

\subsection{The Architectural Challenge}

While the above transition is primarily a technological problem — post-quantum cryptographic primitives are characterized by attributes that distinguish them from traditional ones — the challenges it presents are also architectural. Due to the large difference in artifact size, e.g., ML-DSA-65 public keys are 1,952 bytes while Ed25519 is 32 bytes; ML-KEM-768 ciphertexts are 1,088 bytes vs. 32 bytes for an X25519 key share; and SLH-DSA-SHA2-128s signatures are 7,856 bytes while Ed25519 is 64 bytes, the differences in artifact size affect how protocols are designed, how data is stored, and how network assumptions are impacted. Additionally, there are no straightforward algorithm substitution techniques available to address these issues.

 Most deployed systems embed cryptographic algorithm choices in application source code. Therefore, migrating cryptographic algorithms will remain a major software development task with the potential impact on numerous source files. Lazar et al. \cite{lazar2014} showed that 83\% of the cryptographic vulnerabilities they identified in an analysis of 269 cryptography-related CVEs resulted from misuse of cryptographic APIs, indicating that the relationship between application code and cryptographic implementation has become a pervasive source of problems. Similarly, Acar et al. \cite{acar2017} provided evidence that API design significantly influences the security of the resulting code. Thus, the complexity of the post-quantum transition is not incidental but inherent, as the APIs applications use to access cryptography involve explicit specification of algorithms and properties that are not easy to migrate.

 Prior experience with other cryptographic transitions indicates that similar difficulties exist. For example, the transition from SHA-1 to SHA-256 took approximately ten years from announcement of SHA-1 deprecation until widespread adoption of SHA-256, although the two hash functions have identical operational interfaces (i.e., they both take a single input and produce a fixed-length digest) except for their output lengths. The transition from DES to AES also spanned many years. Both times the main bottleneck was not algorithm availability but instead the amount of software engineering necessary to identify all applications referencing the deprecated algorithm, modify each application, test each modified application, and finally deploy each modified application. The post-quantum transition creates even greater difficulties since not only must algorithms be replaced but the replacement algorithms are expected to have different properties (e.g., longer keys and/or longer signatures) than their predecessors such that code changes will not be simply substitutions of identifiers but possibly need to involve modification to existing application logic concerning data manipulation
 
\subsection{The Ambiguity of Cryptographic Agility}

Cryptography agility --- generally defined as the ability of a system or protocol to move from one set of cryptographic algorithms to another over time without requiring proportional amounts of code modifications \cite{rfc7696} --- has been cited as one possible approach to addressing some of the problems described above. However, the definition of cryptographic agility is somewhat ambiguous in both academic research \cite{towards_common_understanding} and practical usage \cite{NIST_considerations_agility}. In particular, definitions of cryptographic agility vary depending upon the specific context in which it is being discussed. For example, in contexts involving protocols, cryptographic agility typically means that a protocol or system is able to negotiate updates to supported cipher suites (e.g., TLS 1.3 uses cipher suite negotiation to support updates to supported ciphers). This type of agility is limited to updates affecting protocol-level wire formats and does not necessarily provide any level of support for updating applications. On the other hand, in computing systems, cryptographic agility involves replacing existing cryptographic algorithms with newer ones so long as those replacements do not disrupt current operations \cite{NIST_considerations_agility}. Finally, in organizational contexts, cryptographic agility encompasses policy-based control mechanisms allowing administrators to change encryption policies (and thus implicitly enable changes to underlying cryptographic algorithms) without having developers involved \cite{NIST_considerations_agility}. From an operational perspective, cryptographic agility represents the ability to quickly switch between alternative sets of cryptographic implementations when a vulnerability is detected.

Consequently, conflation of these abilities leads to systems that effectively implement some forms of agility while inadvertently impeding others. Moreover, without a unified nomenclature for defining which types of agility a given system implements, it is impossible for organizations to assess their preparedness for a post-quantum transition; it is impossible for vendors to clearly describe the capabilities of their products; and it is impossible for researchers to systematically evaluate candidate approaches.

\subsection{Contributions}

This paper makes three contributions wrt. agility for cryptographic APIs:

\begin{enumerate}
\item A \textbf{component-based assessment framework} that characterizes cryptographic agility along seven orthogonal dimensions for cryptographic APIs. Three coupling dimensions (measuring what application code requires to know about algorithms and providers), an orthogonally-applied decoupling mechanism (measuring how coupled values are externalized), a governance dimension (measuring who controls the decoupling mechanism), and two agility enablers (measuring actual migration capability). The framework captures non-hierarchical profiles that linear maturity models cannot represent.

\item \textbf{Evaluation of six representative deployed systems} (PKCS\#11, OpenSSL~3.0, JCA, Google Tink, AWS~KMS, and HashiCorp Vault Transit) against the framework. It reveals three pervasive and independent gaps: (i)~operation decoupling remains partial while intent-based key creation is absent in all six systems; (ii)~two systems achieve role-based access control, but none provides authority over cryptographic algorithm selection or control. Cryptographic governance is a dimension that is not captured by any of the evaluated system; and (iii)~no system supports explicit algorithm transformation as an independent operation. So even when decoupling exists it cannot be exercised for cross-algorithm migration.

\item \textbf{Derived design requirements} that identify the architectural provisions necessary to close the gaps revealed by the evaluation, providing concrete guidance for API designers and a bridge to the companion paper~\cite{paper2-api} on API design patterns.
\end{enumerate}

\subsection{Paper Organization}

The remainder of this paper is organized as follows. Section~\ref{sec:background} reviews the evolution of cryptographic APIs. Section~\ref{sec:framework} presents the assessment framework. Section~\ref{sec:evaluation} evaluates six deployed systems against the framework and identifies three pervasive gaps. Section~\ref{sec:requirements} derives design requirements from the framework's gap analysis and Section~\ref{sec:related} reviews the related work.

\section{Background and Motivation}
\label{sec:background}

{
\emergencystretch=1em
Cryptographic APIs have evolved through three identifiable generations. Understanding this evolution is essential for identifying the precise gaps that the post-quantum transition needs. Each of those generation achieves partial agility along specific dimensions and resolves certain coupling problems while leaving others untouched. The persistent residual coupling in these APIs motivate the systematic framework we present in Section~\ref{sec:framework}.\par
}

\subsection{First Generation: Algorithm-Specific APIs}

The earliest cryptographic APIs exposed a separate function for each algorithm. OpenSSL's legacy interface shows this pattern: \texttt{RSA\_public\_encrypt()} and \texttt{AES\_cbc\_encrypt()} are distinct entry points that accept algorithm-specific parameter structures. Each algorithm requires its own function, its own parameter types, and often its own key representation. This bakes in the algorithm's identity at every call site.

To understand the severity of the coupling, consider a concrete migration scenario. An application that signs audit log entries with RSA-PSS must specify not only the algorithm but also the hash function, salt length, and encoding format at every call:
\begin{lstlisting}
// First-generation pattern (Go, illustrative)
sig, err := rsa.SignPSS(
    rand.Reader,           // Source of randomness
    privateKey,            // Key material
    crypto.SHA256,         // Hash function (hardcoded)
    hashed[:],             // Pre-hashed data
    &rsa.PSSOptions{       // Algorithm-specific options
        SaltLength: rsa.PSSSaltLengthEqualsHash,
    },
)
\end{lstlisting}
Six decisions are embedded in a single function call: algorithm (RSA-PSS), hash function (SHA-256), key material handling, salt length, randomness source, and signature encoding. Changing \textit{any} of these decisions requires modifying the source code, recompiling, testing, and redeploying. Migrating from RSA-PSS to ECDSA requires restructuring the parameter list entirely, because ECDSA accepts different options (curve selection). Migrating to a post-quantum algorithm such as ML-DSA would require yet another restructuring, since ML-DSA introduces a domain-separation context parameter that RSA-PSS does not accept.

First-generation APIs thus exhibit maximal coupling along three distinct dimensions. Every call site encodes the algorithm's identity and parameters, so operations cannot be reused across algorithms. Key generation requires explicit algorithm and parameter specification, so key creation sites must change when the algorithm changes. The cryptographic implementation is an import-time dependency, so changing the library also requires modifying the source code. The migration cost scales as $O(n \cdot m)$, where $n$ is the number of call sites and $m$ is the number of parameter differences between the old and new algorithms.

\subsection{Second Generation: Operation-Level Abstraction}

{\emergencystretch=2em
The second generation APIs introduced the concept of \textit{uniform operation interfaces} that abstract away algorithm-specific function names for cryptographic operation. PKCS\#11~\cite{pkcs11} provides generic functions such as \texttt{C\_Sign()}, \texttt{C\_Encrypt()}, and \texttt{C\_Decrypt()} that apply across all algorithms. The Java Cryptography Architecture (JCA)~\cite{jca1997} adopted a similar model, providing factory methods such as \texttt{Signature.get\-Instance("SHA256\-withRSA")} that return a common \texttt{Signature} object. The OpenSSL~3.0 EVP interface~\cite{openssl-evp} provides \texttt{EVP\_Digest\-Sign*()} and \texttt{EVP\_Digest\-Verify*()} functions that accept algorithm identifiers as parameters. Google's Tink~\cite{tink2018} represents the most advanced expression of this generation, defining primitive-based interfaces. texttt{Aead}, \texttt{Public\-Key\-Sign}, \texttt{Hybrid\-Encrypt} group algorithms by their cryptographic function rather than merely unifying function signatures.\par}

{\emergencystretch=2em
A uniform function signature does not by itself produce algorithm-agnostic operations. The degree of decoupling depends on what information the caller must supply beyond the function name. In PKCS\#11, \texttt{C\_Encrypt()} requires a \texttt{CK\_MECHANISM} field whose parameter structure varies per algorithm. For example, AES-GCM demands a \texttt{CK\_GCM\_PARAMS} structure carrying IV, AAD, and tag-length fields. Similarly, RSA-OAEP requires a \texttt{CK\_RSA\_PKCS\_OAEP\_PARAMS} structure specifying the hash function, mask generation function, and source data~\cite{pkcs11}. JCA exhibits the same pattern through its \texttt{Algorithm\-Parameter\-Spec} hierarchy. OpenSSL's EVP interface exposes algorithm-specific details through its \texttt{EVP\_CIPHER\_CTX\_ctrl()} mechanism. Tink is the notable exception in this list. Its \texttt{Aead.encrypt(plaintext, associated\-Data)} interface is truly algorithm-agnostic, demonstrating that complete operation-level decoupling is architecturally achievable.\par}

Despite this variation in decoupling for cryptographic operations, second-generation systems universally preserve algorithm coupling at key creation. The algorithm must still be specified at the point of key generation. PKCS\#11 requires a \texttt{CK\_MECHANISM} specifying the algorithm in \texttt{C\_GenerateKeyPair()}, JCA requires \texttt{KeyPairGenerator.getInstance("RSA")} and Tink locks the algorithm at keyset creation through its \texttt{KeyTemplate} mechanism. This residual coupling at key creation has two consequences. First, any change in the algorithm still requires modifying key creation code. Second, the asymmetry between the expectation during cryptographic operation and key creation produces an \textit{incomplete agility profile}. Cryptographic operations appear interchangeable, but the key creation sites still encode algorithm-specific knowledge, so the system as a whole cannot migrate without code changes.

In summary, second-generation systems partially decouple operations from algorithms. Function signatures are uniform, but algorithm-specific knowledge leaks through parameter structures in PKCS\#11, JCA, and OpenSSL EVP. Only Tink achieves full operation-level decoupling. However, key creation remains fully coupled to specific algorithms across all four systems, and provider abstraction varies. The net effect is an asymmetry between operations and creation that the assessment framework of Section~\ref{sec:framework} will characterize precisely.

\subsection{Third Generation: Governance-Aware Key Management}

Cloud key management services such as AWS KMS, Google Cloud KMS, Azure Key Vault, and HashiCorp Vault Transit add a governance layer that previous generations lacked. These systems provide key storage, access control through IAM policies, audit logging, and hardware security module (HSM) backing for protecting keys and cryptographic operation. Hosting keys in a managed service eliminate the need for applications to handle raw key material. We evaluate AWS~KMS and Vault Transit in the Section~\ref{sec:evaluation}

Despite the advances along the governance dimension, algorithm coupling exists even in this generation of APIs. AWS KMS requires \texttt{KeySpec: "RSA\_2048"} at key creation. HashiCorp Vault Transit requires \texttt{type: "ecdsa-p256"} when creating a named key. In each case, the application must specify the concrete algorithm at key creation time, and none of these services supports changing the algorithm of an existing key. Vault Transit provides rotation (\texttt{rotate/:name}) of keys under the same algorithm, but not cross-algorithm transformation.

The governance mechanisms in third-generation systems address the access control aspects but not cryptographic control. For example, an IAM policy can restrict key creation to authorized principals, but no existing governance mechanism can express a constraint about which cryptographic algorithms are allowed, and should be used. This distinction, between access authority and cryptographic selection authority, is a key aspect our assessment framework will formalize.

Therefore, third-generation systems advance some abstraction and access control, but algorithm coupling at key creation still persists. Algorithm migration is also restricted to same-algorithm rotation.

\subsection{The Persistent Gap}

Each generation of cryptographic API resolves one category of coupling while leaving others intact. Coupling between cryptographic operations and specific algorithms are substantially addressed by second- and third-generation systems. Third-generation cloud services address the access control governance requirements. But coupling at key creation ( the need to specify concrete algorithms when keys are generated) remains universal across all the three generations. The ability to migrate algorithm for a key is also absent across all the deployed systems we surveyed.

These gaps are individually sufficient to prevent agile migration. This means every key creation site must be modified to adopt a new algorithm. Without the ability to change algorithms, every existing key must be abandoned and recreated, and all data protected under old keys must be re-processed. Together, they increase the burden of transitioning between algorithms and make it a large software development project. The assessment framework presented in Section~\ref{sec:framework} provides the analytical vocabulary necessary to characterize these gaps precisely.

\section{The Cryptographic Agility Assessment Framework}
\label{sec:framework}

\begin{figure*}[t]
\centering
\includegraphics[width=0.82\linewidth]{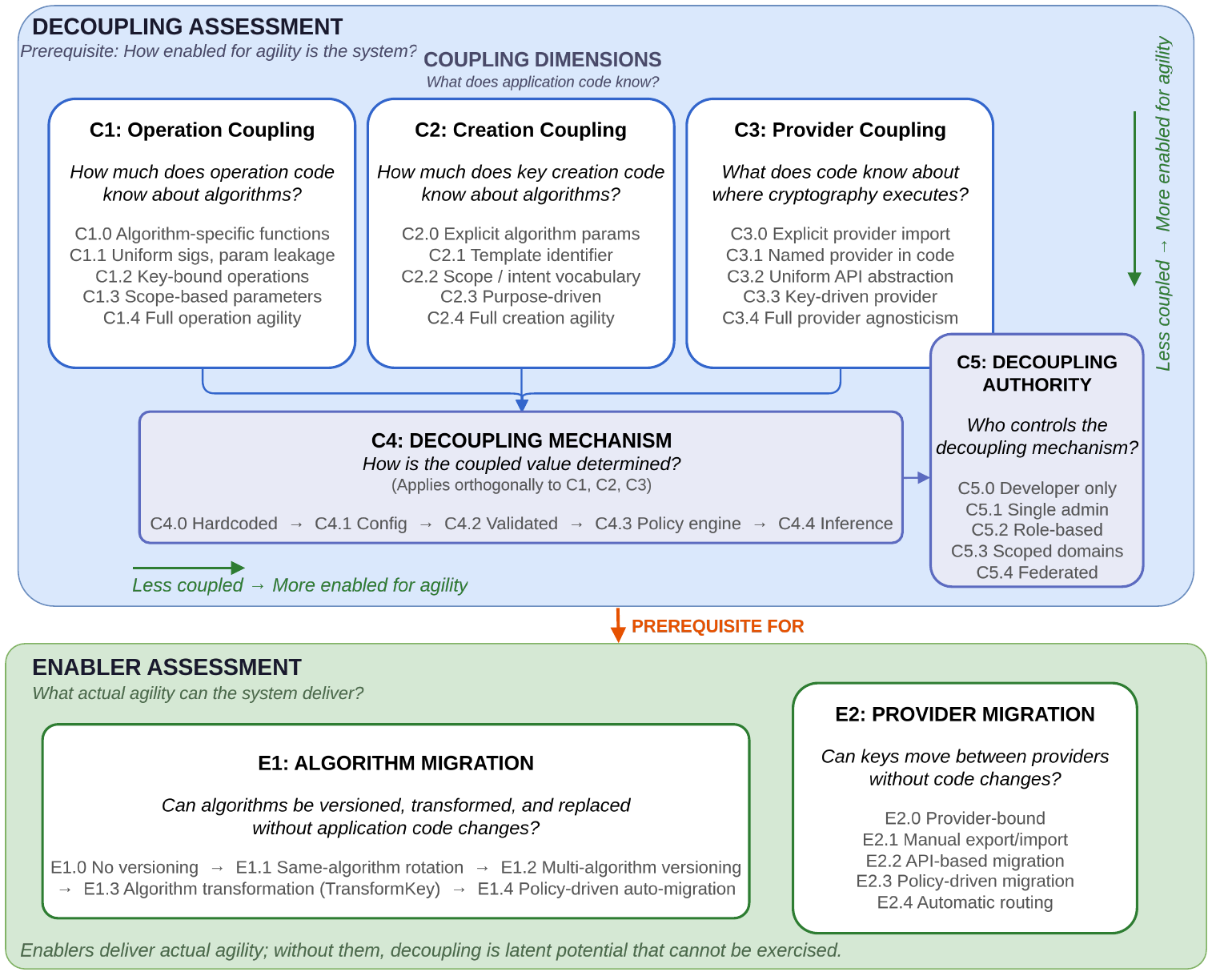}
\caption{The Cryptographic Agility Assessment Framework: seven orthogonal components organized into coupling dimensions (C1--C3), decoupling mechanism (C4), decoupling authority (C5), and agility enablers (E1, E2).}
\label{fig:framework}
\end{figure*}

We present a component-based assessment framework (Figure~\ref{fig:framework}) for cryptographic APIs. It characterizes cryptographic agility along orthogonal dimensions rather than a single linear hierarchy. The framework distinguishes between \textit{decoupling}, which creates the conditions under which agility becomes possible, and \textit{agility enablers}, which deliver the capability for actual migration. Without agility enablers, the system may be highly decoupled but is still unable to migrate existing keys to new algorithms.

\subsection{Framework Architecture}

The framework comprises seven components organized into two assessment tiers.

\noindent\textbf{Tier 1: Decoupling Assessment.} Five components measure how \textit{enabled} for agility a cryptographic API is. Three \textit{coupling dimensions} measure what the application code knows about specific values: C1 (operation coupling, measures algorithm awareness required during cryptographic operations), C2 (creation coupling, measures algorithm awareness required during key creation), and C3 (provider coupling, measures awareness of where cryptography executes). C4 (decoupling mechanism) measures how coupled values are externalized, and applies orthogonally to each coupling dimension. C5 (authority) measures who controls the decoupling mechanism. 

\noindent\textbf{Tier 2: Enabler Assessment.} Two components (E1, E2) measure the \textit{actual agility} the system can deliver. E1 (algorithm migration) captures the ability to version, transform, and replace algorithms. E2 (provider migration) captures the ability to move keys between providers. These are orthogonal to decoupling but mutually reinforcing. High decoupling without enablers produces unused potential that cannot be exercised, and enablers without decoupling produce ecosystem-locked agility that cannot be generalized.

\noindent\textbf{Level Notation.} Each component is assessed on a spectrum from 0 (maximal coupling or no capability) to 4 (maximal decoupling or full capability), denoted by C$n$.$x$ or E$n$.$x$. On first introduction, each level is accompanied by a descriptive label (e.g., C2.2, \textit{intent-based creation}); subsequent references use the compact notation.

\subsection{Coupling Dimensions}

\paragraph{C1: Operation Coupling.} 
This dimension measures the extent to which cryptographic operation call sites are coupled to specific algorithms. The various levels are determined by the category of information leaking through the operation’s parameters.

\noindent\textbf{C1.0, Algorithm-specific functions. } Each algorithm requires its own function: \texttt{RSA\_sign()}, \texttt{ECDSA\_sign()}, and \texttt{ML\_DSA\_sign()} each represent separate entry points. Since the function name itself represents the algorithm used, changing the algorithm requires updating all calls to a new, structurally distinct function. Thus, this cost increases linearly with respect to the total number of operations within a code base.

\noindent\textbf{C1.1, Uniform signature with parameter leakage.} The improvement over C1.0 is that the function name is no longer an indication of the algorithm. For example, using a single \texttt{sign()} can replace both \texttt{RSA\_sign()} and \texttt{ECDSA\_sign()}. However, the distinction between the algorithms is moved from the function name to the function’s parameters. Therefore, the caller must provide some values whose structure or semantic content will differ depending upon the algorithm utilized. There are three types of values: (i) algorithm identity (name or identifier of mechanism), (ii) system-generatable values whose format differs among algorithms (IV, IV-size, tag-length, salt-length), and (iii) user-provided operational parameters (AAD, domain separation context) that only the application can provide during the cryptographic operation. Although the type of coupling remains similar, it now exists in a different form: all three categories must be updated if the algorithm changes.

\begin{sloppypar}

\noindent\textbf{C1.2, Dispatching based on keys.} The improvement over C1.1 is the elimination of two of the three categories of algorithm-specific information from the call site. Categories (i) and (ii) (algorithm identity and system generatable values) are incorporated into key metadata. The system determines which algorithm to use based on the key's identity and internally generates values such as IVs and tag lengths from the key's stored configuration. Calls to cryptographic primitives are dispatched based on key identity rather than algorithm identity. The caller provides a reference to the key along with user data, and the system determines all other necessary internal parameters for encryption/decryption. Category (iii) (user-provided operational parameters such as AAD for AEAD or context for domain separation) may also exist at the point where calls are made, and those parameters may still vary between algorithms.

\end{sloppypar}

\begin{sloppypar}

\noindent\textbf{C1.3, Scope consistent operations.} The advancement from C1.2 is to remove differences in user-supplied operational parameters across algorithms within the same cryptographic primitive. In C1.2, different algorithms within the same primitive may take different user-supplied parameters. For example, one signature algorithm may expect an explicit context while another signature algorithm cannot take a context as input. C1.3 requires that all algorithms within a given scope will accept the same set of user-supplied parameters with the same semantic definitions. The constraint imposed upon user-supplied operational parameters by C1.3 is that they must be \textit{scope consistent}. With this constraint, true algorithm substitutability holds. Therefore, calling one algorithm instead of another within the same scope would require no modifications to existing call sites, since the interface for passing parameters to operations would be identical.

\end{sloppypar}

\begin{sloppypar}
\noindent\textbf{C1.4, Fully managed operations.} The advance from C1.3 is the elimination of user-supplied operational parameters entirely. Where C1.3 permits scope-consistent parameters (e.g., context for signatures), C1.4 removes even these from the call site: the application provides only the plaintext, ciphertext, or message to be processed. The system manages all cryptographic parameters internally and achieves maximum substitutability at the cost of application flexibility.
\end{sloppypar}

\smallskip

\noindent A major trade-off in terms of \textit{agility and flexibility} exists throughout this spectrum. Applications operating at the highest level of agility will achieve such by either removing user choice, or by forcing users to choose. As a result, applications requiring domain-specific context or IVs, etc., may not be possible at this level and/or provide the full flexibility to support all possible use cases with user-defined input. C1.3 on the other hand, provides a balanced solution. C1.3 maintains substitution via the fact that all operations of the cryptographic API can take on the same set of user-provided parameters, and therefore supports the required flexibility for real-world applications.

\paragraph{C2: Creation Coupling.}
Creation coupling refers to how tightly the knowledge of the algorithms used are tied to the process of creating keys. Higher levels (C2.0 through C2.4) indicate greater degrees of decoupling.

\noindent\textbf{C2.0, Raw Algorithm Parameters.} At this level of creation coupling, the application defines both the algorithm and its associated parameters when creating keys. For instance, (\texttt{algorithm=\allowbreak{}"RSA",\allowbreak{} key\_size=\allowbreak{}2048}) needs to be passed into the key-creation method. Therefore if you decide to change your chosen algorithms you would have to modify each point where keys were created.

\begin{sloppypar} 
\noindent\textbf{C2.1, Named Template Identifiers.} 
Named Template Identifiers is one step above C2.0 and adds an additional abstraction layer to avoid listing raw parameters, instead allowing an application to identify the algorithm and parameters using named identifiers. Thus an application does not need to list out the individual parameters (e.g. {algorithm, key size, mode, hash function, padding}), but rather specify something like {\texttt{template=\allowbreak{}"rsa-2048-oaep-sha256"}}. Therefore, there are fewer coupled values: specifically down from many parameters (algorithm, key size, mode, hash function, padding) to a single string identifier. The identification of the template makes it easier to create keys, while also acting as a surrogate for the selection of a specific algorithm. Similar to C2.0, an application requesting {\texttt{template=\allowbreak{}"rsa-2048-oaep-sha256"}} has made a definitive choice regarding what algorithm(s) they wish to employ. If you replace that algorithm you'll again have to make that new algorithm known to each location where keys are being created.
\end{sloppypar}

\noindent\textbf{C2.2, Intent-based creation.} This level represents the qualitative shift in abstraction. Instead of specifying a name for an algorithm by the application, the application specifies its own crypto-logical intent via a vocabulary (i.e. \texttt{scope=\allowbreak{}AUTHENTICATED\_ENCRYPTION}). It only indicates which security properties it wishes to achieve using the vocabulary during the process of creating keys. The system will determine the most suitable algorithm on behalf of the application based on either a policy or configuration. Unlike a named template which specifies a concrete algorithm and a set of parameters, the intent vocabulary can map to multiple templates that meet the security properties of the intent.

\noindent\textbf{C2.3, Purpose-driven requests.} Compared with C2.2, C2.3 advances by removing the reference made within the application to the specific crypto-concepts of authenticated encryption or digital signature, and expressing business-level intent (i.e. \texttt{purpose=\allowbreak{}"sign audit logs"}) instead. In other words, the mapping from business purposes to crypto-scopes is determined outside of the application. Therefore, if there were a change in how a particular business purpose was interpreted cryptographically (e.g. "sign audit logs" from being interpreted as digital signatures to authenticated MACs), then the application itself does not need to be modified. At this point, application developers no longer need to know anything about primitive cryptographic constructs. However, this also means that the cryptographic API is custom-designed or enabled for multiple purpose-driven use cases.

\noindent\textbf{C2.4. Full contextual inference.} Compared to C2.3, C2.4 involves the removal of all explicit specifications from the application. That is, the system will infer the proper cryptographic configuration based solely on environmental context. The environmental context can include many types of information including: data classification, regulatory jurisdiction, communication protocols, threat models, etc., and may include none of these categories of information and/or may not even require a purpose to be declared. Therefore, C2.4 represents the ultimate separation of the application code from all crypto-related decision-making processes. However, this can only occur when the system has enough knowledge about environmental context to make autonomous decisions.

\smallskip
\noindent The primary design challenge associated with C2.2 is \textit{designing scope boundaries.} Algorithm scopes must be defined so that they contain algorithms whose operational parameters supplied by users during crypto-related operation are identical in order to avoid breaking the application code when substitution occurs. If too wide a scope is selected for an algorithm, then substitution hazards exist, whereas selecting too narrow a scope eliminates the benefits of abstraction.

\paragraph{C3: Provider Coupling.}
This dimension measures the degree to which application code is coupled to a specific cryptographic implementation or execution environment.

\noindent\textbf{C3.0, Compile-time dependency.} The provider is an explicit compile-time or link-time dependency. Switching providers requires modifying the source code or build configuration.

{\emergencystretch=3em
\noindent\textbf{C3.1, Named provider selection.} The advance from C3.0 is that a uniform framework mediates access to multiple providers through a common interface, but the application still explicitly selects a provider by name at the call site (e.g., JCA's \texttt{Signature.get\-Instance("RSA",~"BC")}). At this level, the provider dependency has moved from a compile-time import to a runtime parameter. The provider identity still remains in the application source code. Switching providers requires modifying the provider name at each call site and not the surrounding API surface.\par}

\begin{sloppypar}
\noindent\textbf{C3.2, Uniform API abstraction.} The advance from C3.1 is the removal of provider identity from application code entirely. The same interface functions identically regardless of which backend executes the operation. At C3.1, the application names the desired provider, and at C3.2, the application makes no provider-related assertion. Provider selection is determined by external system configuration, so an administrator can redirect all operations to a different backend without any application modification.
\end{sloppypar}

\begin{sloppypar}
\noindent\textbf{C3.3, Key-driven provider selection.} The advance from C3.2 is that provider selection becomes data-driven rather than configuration-driven. At C3.2, a system-wide configuration determines the provider for all operations. At C3.3, the provider is resolved individually for each key based on where the key is stored. An application that references a key stored in an HSM automatically receives HSM-backed operations, and a key in software is serviced by a software provider, all without any provider-specific code. This enables multiplexing providers within a single application.
\end{sloppypar}

\noindent\textbf{C3.4, Full provider agnosticism.} The advance from C3.3 is that dynamic provider routing becomes possible. The system manages all provider routing decisions transparently. Aspects like load balancing, failover, capability-based dispatch, and compliance-aware routing are possible at this level. Constraint-based provider use, such as selecting and reselecting providers based on real-time availability, latency, and regulatory requirements, falls in this level.

\subsection{Decoupling Mechanism and Authority}

\subsubsection{C4: Decoupling Mechanism.}
\textit{How is the coupled value determined?} C4 measures how the coupled values are externalized, and applies independently to each coupling dimension (C1, C2, C3). This dimension is meaningful only when the C1, C2, or C3 components are at a level that has coupled values.  

\noindent\textbf{C4.0, Hardcoded in source.} Coupled values are hardcoded and expected to be so. Therefore, at this level, any change requires code modification.

\begin{sloppypar}
\noindent\textbf{C4.1, Configuration file.} The advance from C4.0 is the separation of the coupled value from source code. An algorithm identifier, provider name, or key parameter that was previously embedded in application code is now read from a configuration file, environment variable, or similar external source. This permits changes without recompilation or redeployment. However, the configuration content is syntactically and semantically unconstrained. Therefore, invalid, insecure, or contradictory values are accepted without detection.
\end{sloppypar}

\begin{sloppypar}
\noindent\textbf{C4.2, Schema-validated configuration.} The advance from C4.1 is the introduction of validation on the externalized values. At this level, a schema constrains the space of permissible configurations. For example, type checking to ensure correct values are used, enumeration constraints to restrict values to known algorithms, and structural rules to enforce required fields and valid combinations are aspects of this level. Schema validation operates only at the syntactic level and cannot express semantic or conditional constraints, such as ``this algorithm is permitted only in production environments.''
\end{sloppypar}

\noindent\textbf{C4.3, Policy engine.} At this level, rules-based mapping not only enforces syntactic requirements but also allows for semantic validation, conditional logic, and audit trails. This level introduces the possibility of cryptographic governance and control, which C4.2 does not support. A policy can express constraints such as ``use only FIPS-approved algorithms with 256-bit security for production environments.''

{\emergencystretch=2em

\noindent\textbf{C4.4, Full contextual inference.} Moving forward from C4.3 means eliminating the specification of an explicit configuration altogether. The system will determine the most suitable cryptography from the context of the environment. These context parameters can include; e.g. data sensitivity, legal jurisdictions for the regulation, the relevant threat model, etc. without any administrator having to explicitly state them, so they are inferred correctly by the system autonomously.

\smallskip
\noindent C4 applies orthogonally to C1, C2, and C3. For example, a system at C1.1 may apply C4.0 to C1, C4.3 to C2, and C4.1 to C3, each of which is assessed independently. However, C4 is only applicable to a coupled dimension when that dimension has a value that needs to be determined. If a coupled dimension is completely resolved (e.g. at C1.2+; where the metadata of the key dictates the choice of the algorithm), then C4 is inapplicable to this dimension since there is no value for the externalization mechanism to control.

\subsubsection{C5: Decoupling Authority.}
\textit{Who controls the decoupling mechanism?} C5 measures the degree of support from the cryptographic API and the granularity of an authority to enforce the decoupling mechanism in C4.

\noindent\textbf{C5.0, Developer-only authority.} At this level, all cryptographic decisions are made by the developer, and they are the sole authority.

\noindent\textbf{C5.1, Single administrator.} The advance from C5.0 is the separation of control from development. A designated administrator can modify the cryptographic configuration without requiring code changes. However, this administrator also possesses undifferentiated authority over other functions. The same principal who selects algorithms also configures providers and defines lifecycle constraints, with no separation of duties.

\noindent\textbf{C5.2, Role-based authority.} At C5.2, different actors can control different aspects of configuration. For example, a security architect defines approved algorithms, a compliance officer enforces FIPS requirements, and a developer selects from the approved set of algorithms.

\noindent\textbf{C5.3, Scoped governance domains.} The advance from C5.2 is the introduction of \textit{governance domains}. The ability to have distinct organizational contexts (e.g., business units, application tiers, geographic regions) maintain independent cryptographic policies managed by separate authorities. For example, a payment processing domain may enforce FIPS-only algorithms at 256-bit security, while a development sandbox permits experimental post-quantum algorithms. Each domain authority operates independently, and the system can enforce domain boundaries.

\noindent\textbf{C5.4, Federated governance.} The advance from C5.3 is the introduction of policy composition across governance domains. A corporate authority defines baseline constraints (e.g., a minimum 128-bit security requirement across all domains), and domain-level authorities extend but cannot weaken these baselines. Policy inheritance is possible, ensuring consistency at the organizational level while permitting local specialization.

\smallskip
\noindent C5 is meaningful only when C4 $\geq$ C4.1. Crucially, C5.2 (role-based \textit{access} authority) is distinct from C5.3+ (authority over \textit{cryptographic selection}). This distinction is examined in the system evaluations of Section~\ref{sec:evaluation}.

\subsection{Agility Enablers}

Decoupling determines how \textit{able} a system is to achieve agility. Agility Enablers determine if you can actually achieve it. Without agility enablers, even if your systems were perfectly decoupled they would still not be able to switch their keys to other algorithms.

\subsubsection{E1: Algorithm Migration.}
\textit{E1 measures whether algorithms can be versioned, modified, and replaced without requiring any modifications to the application’s source code.}

\noindent\textbf{E1.0, No versioning.} At this level, the key identity is permanently tied to the algorithm that generated the key. Therefore, when you want to adopt a new algorithm, you have to create a new key with a new key identifier. Every reference to the key in every consumer has to be updated. The cost to move from one algorithm to another is directly proportional to the number of consumer applications that used the old key.

\noindent\textbf{E1.1, Same-algorithm rotation.} The advance from E1.0 is the introduction of key versioning within a fixed algorithm. The system generates new key material under the same algorithm while preserving the logical key identifier. New operations use the latest version, while previously protected data is processed using the version active at the time of protection. This supports periodic key refresh and limits exposure of individual key versions, but cannot change the underlying algorithm. For example, an AES-256 key can only be rotated to another AES-256 key.

\noindent\textbf{E1.2, Multi-algorithm versioning.} The next step after E1.1 is to separate the identity of the key from the identity of the algorithm on which the key is based across multiple versions. Thus, one logical key may include versions created using different algorithms. For example, version~1 of the key is created using ECDSA, while version~2 of the key is created using ML-DSA. As new operations are executed, the system will select and use the most recent version of the key. If older data needs to be processed, the correct version of the key valid at the time of processing will be selected. This allows incremental migrations from one algorithm to another while ensuring that access to historical data remains available. However, no specific operation is defined to distinguish between simply replacing the key's bytes under the same algorithm and switching to a completely new algorithm.

\noindent\textbf{E1.3, Explicit algorithm transformation.} The next step after E1.2 is for the system to provide an explicit first-class operation to perform migrations between algorithms. Rather than treating a simple rotation of a key's bytes with the same algorithm the same as migrating between algorithms under that same key, an explicit transformation operation (for example, TransformKey()) creates a new set of key material under the provided target algorithm. Also, it will update the current key version and record the transformation in the version history. Thus, making transformations between algorithms a policy-governed, auditable operation, amenable to access control rather than an ad hoc one.

\noindent\textbf{E1.4, Policy-driven automatic migration.} The advance from E1.3 is the automation of algorithm migration. When a policy change alters the required or preferred algorithm, the system automatically identifies affected keys and triggers transformation operations without administrator intervention. At this level, decisions are declarative, and no imperative commands are required for migration. Organization-wide algorithm transitions reduce from per-key administrative procedures to a single policy update.

\smallskip
{\emergencystretch=1em
\noindent The distinction between E1.1 and E1.3 is critical for the post-quantum transition: same-algorithm rotation cannot transform an RSA-2048 key into an ML-DSA-65 key.\par}

\subsubsection{E2: Provider Migration.}
\textit{E2 assesses the ability to migrate keys and cryptographic operations between providers (i.e., execution environments) with no modifications to the application code.}

\noindent\textbf{E2.0, Provider-bound keys.} Keys are always permanently associated with the provider that generated the keys. Therefore, any time you wish to switch providers, you have to create new keys within the target provider, protect all of your data under those new keys and modify all of the references to keys. This represents the same type of cost as E1.0 for migrating algorithms.

\noindent\textbf{E2.1, Manual Export/Import.} The step up from E2.0 is that you can extract key material from one provider and import into another via some form of out-of-band manual process (e.g., \texttt{C\_WrapKey}/\texttt{C\_UnwrapKey} or a KMS export/import workflow.) In order to perform this task you need to actively engage administration and manually handle the exported key material. Whether the key ID will be maintained or not is dependent upon how you implement this process and the lack of any transactional guarantees is inherent.

\noindent\textbf{E2.2, Explicit Migration API.} The step up from E2.1 is the presence of a first class migration API that coordinates all aspects of the migration as a single atomic operation. The system takes care of extracting key material, rewrapping under protection key for target provider, coordinating metadata updates and preserving identifiers atomically. By doing so, you eliminate manual handling of raw key material and potentially gain transactional guarantees.

\noindent\textbf{E2.3, Policy driven migration.} The advancement over E2.2 is automating provider migrations using policy based management. As policies evolve and indicate a preferred provider (i.e., due to changes in preference for HSM backed keys for production workloads) -- the system will automatically identify impacted keys and initiate migration operations. This reduces provider migrations from an administrative effort per key to simply changing policies.

\noindent\textbf{E2.4, Automatic provider routing.} The advancement from E2.3 is that provider selection becomes continuously dynamic rather than being triggered by policy changes. The system selects providers for each operation based on current contextual information (e.g. latency, availability, jurisdiction for compliance, cost, etc.) and will dynamically route or migrate key material as required to support the current constraints.

\smallskip
\noindent There are legitimate security concerns that constrain provider migration: HSM-bound keys cannot be exported by definition, therefore the best E2 level is dependent on your security model.

\subsection{Constraints and Post-Quantum Implications}

\subsubsection{Structural Constraints.}
Six structural constraints govern valid combinations of components, defining the dependency relationships among them.

\begin{enumerate}
\item \textbf{Externalization requires parameterization.} Applying C4.1+ to operations requires at least C1.1.

\item \textbf{Provider externalization requires provider selection.} Applying C4.1+ to providers requires at least C3.1.

\begin{sloppypar}
\item \textbf{Policy requires a scope vocabulary.} Policy-based algorithm selection (C4.3+ on C2) requires a vocabulary~(C2.2+) for expressing algorithm constraints.
\end{sloppypar}

\begin{sloppypar}
\item \textbf{Governance requires externalization.} Governance~(C5.2+) requires externalization of values on at least one coupling dimension~(C4.1+).
\end{sloppypar}

\begin{sloppypar}
\item \textbf{Algorithm transformation benefits from policy.} While E1.3 can function independently, it is most effective combined with C4.3+ on~C2, enabling the policy engine to determine the target algorithm automatically.
\end{sloppypar}

\begin{sloppypar}
\item \textbf{Provider migration is security-model-constrained.} E2.2+ is constrained by the security model of the providers involved.
\end{sloppypar}
\end{enumerate}

\subsubsection{Assessment for Post-Quantum Readiness.}
Our framework quantitatively assesses post-quantum readiness as the cost of algorithm migration as a function of the components. For example, consider the concrete case of migrating from ECDSA P-256 to ML-DSA-65.

For a configuration with minimal agility (C2.0, c4.0, E1.0), migrating to ML-DSA-65 requires: (1)~finding where every key creation was called, (2)~changing those calls to include ML-DSA-65, (3)~adjusting the operational parameters for ML-DSA (since ML-DSA includes an optional context parameter), (4)~creating new keys; (5)~re-signing all existing data, and (6)~retiring the old keys. With many applications, this could take a significant amount of time.

In contrast, with the most agile configuration (C2.2, C4.3, E1.3) then the very same migration will reduce to: (1)~update the security policies, (2)~transform keys using \texttt{transformkey()}, and (3)~all new operations will automatically use the current version of everything. No need to modify any application code, and previously signed data continue to be verifiable, giving enough time to re-sign them. 

\section{Evaluation: System Assessment Against the Framework}
\label{sec:evaluation}

We apply the assessment framework to six representative cryptographic systems, spanning the three API generations identified in Section~\ref{sec:background}: libraries (PKCS\#11, OpenSSL~3.0, Tink), frameworks (JCA), and managed services (AWS KMS, HashiCorp Vault Transit). This selection was guided by three criteria: coverage of the three generational paradigms, diversity of architectural approaches, and prevalence in enterprise deployment.

For each system, we assigned component positions by examining the public API surface, official documentation, and source code where available. The assessment criteria are the level definitions from Section~\ref{sec:framework}: each system is placed at the highest level that its conditions \textit{fully} satisfy. We evaluate systems' standalone capabilities without accounting for what consumers might abstract away. Table~\ref{tab:assessment} summarizes the results.

\begin{table*}[t]
\caption{System assessment: qualitative agility profiles. \textsuperscript{\dag}Value read from configuration rather than hardcoded (C4.1). \textit{Operations}~(C1): Uniform sig (C1.1) = shared function signature but algorithm-specific parameters leak; Key-dispatch (C1.2) = dispatch on key metadata with no algorithm parameters; Fully managed (C1.4) = system handles all parameters. \textit{Key Creation}~(C2): Raw params (C2.0) = explicit algorithm and parameters; Named tmpl (C2.1) = named template identifiers. \textit{Provider}~(C3): Hardcoded (C3.0) = compile-time dependency; Named (C3.1) = named provider selection; Uniform API (C3.2) = provider-agnostic interface; Key-driven (C3.3) = provider determined by where key resides. \textsuperscript{\ddag}Enterprise only. \textit{Governance}~(C5): Access ctrl (C5.2) = role-based authority. \textit{Algo.\ Migr.}~(E1): Same-algo (E1.1) = same-algorithm rotation; Multi-algo (E1.2) = multi-algorithm versioning. \textit{Prov.\ Migr.}~(E2): Manual exp.\ (E2.1) = manual export/import.}
\label{tab:assessment}
\centering
\footnotesize
\setlength{\tabcolsep}{3pt}
\resizebox{\linewidth}{!}{%
\begin{tabular}{@{}lllllll@{}}
\toprule
\textbf{System} & \textbf{Operations} & \textbf{Key Creation} & \textbf{Provider} & \textbf{Governance} & \textbf{Algo.\ Migr.} & \textbf{Prov.\ Migr.} \\
\midrule
PKCS\#11      & Uniform sig     & Raw params    & Uniform API  & None         & None       & Manual exp.  \\
OpenSSL 3.0   & Uniform sig     & Raw params    & Named\textsuperscript{\dag}       & None         & None       & None         \\
JCA           & Uniform sig     & Raw params    & Named         & None         & None       & None         \\
Google Tink   & Fully managed   & Named tmpl    & Hardcoded     & None         & Multi-algo & None         \\
AWS KMS       & Uniform sig     & Named tmpl    & Uniform API\textsuperscript{\dag} & Access ctrl  & Same-algo  & Manual exp.  \\
Vault Transit & Key-dispatch    & Named tmpl    & Key-driven\textsuperscript{\ddag} & Access ctrl  & Same-algo  & Manual exp.  \\
\bottomrule
\end{tabular}}
\end{table*}

\subsection{System-by-System Assessment}

\noindent\textbf{PKCS\#11.} PKCS\#11~\cite{pkcs11} provides uniform function signatures (\texttt{C\_Sign()}, \texttt{C\_Encrypt()}, \texttt{C\_Digest()}) that accept a session handle, data, and a mechanism specification. However, the \texttt{CK\_MECHANISM} parameter structure encodes both algorithm identity and algorithm-specific parameters. Changing from RSA-PSS to ECDSA requires modifying the mechanism and removing the parameter structure. This places operations at C1.1.

Key creation requires explicit mechanism specification (\texttt{C\_GenerateKeyPair()} with algorithm and attributes), constituting C2.0. The slot/token abstraction provides C3.2 (uniform interface across HSM vendors). All coupling dimensions are hardcoded (C4 at 0 on all dimensions). No governance layer exists beyond token-level PIN authentication (C5.0). Migration is limited to manual \texttt{C\_WrapKey}/\texttt{C\_UnwrapKey} sequences (E2.1), and there is no key versioning (E1.0).

{\emergencystretch=1em
\noindent\textbf{OpenSSL 3.0.} The EVP layer provides uniform function signatures, but algorithm-specific values leak through the \texttt{OSSL\_PARAMS} interface: AES-CCM requires the plaintext length to be set before processing the first plaintext block but other AEAD modes do not, valid values for tag/IV length differ across AEAD modes, 
and RSA-PSS requires mode-specific parameters. This places operations at C1.1.\par} The provider architecture in OpenSSL~3.0 loads providers dynamically, with the provider name typically in \texttt{openssl.cnf} (C3.1, C4 at 1 on C3). Key creation remains algorithm-explicit (C2.0, C4 at 0 on C1 and C2). There exists no governance mechanism (C5.0), no key versioning (E1.0), and no provider migration (E2.0).

{\emergencystretch=3em
\noindent\textbf{Java Cryptography Architecture (JCA).} Engine classes (\texttt{Cipher}, \texttt{Signature}, \texttt{Key\-Generator}) provide uniform operation interfaces. However, algorithm names appear as string parameters---\texttt{Cipher.get\-Instance(\linebreak[2]"AES/\allowbreak{}GCM/\allowbreak{}NoPadding")} and \texttt{Signature.get\-Instance("SHA256\-withRSA")}---encoding algorithm identity. This places operations at C1.1.\par}

The JCA provider model supports named provider selection (C3.1). Key creation is algorithm-explicit (C2.0). All coupling dimensions are hardcoded (C4 at 0). There exists no governance, versioning, or provider migration (C5.0, E1.0, E2.0).

\noindent\textbf{Google Tink.} Tink achieves the highest operation decoupling among the evaluated systems. Primitive interfaces (\texttt{Aead}, \texttt{digital signature}) accept only user data. All algorithm-specific values are managed internally (C1.4). This comes at the cost of flexibility, and applications cannot supply custom IVs, specify tag lengths, or provide domain-separation context beyond the associated data parameter for the AEAD primitive.

Key creation uses algorithm-specific template names (\texttt{AES128\_GCM}, \texttt{ECDSA\_P256}), placing creation at C2.1. Keyset management provides multi-algorithm versioning with automatic selection of the primary key (E1.2): this is the most advanced migration mechanism among the evaluated systems. However, Tink's cryptographic implementations are compiled into the library itself, with no mechanism to load alternative providers at runtime, resulting in a compile-time dependency on providers (C3.0). Tink hardcodes all selections (C4.0), provides no external policy (C5.0), and no provider migration (E2.0). Tink achieves agility through \textit{restrictiveness} and \textit{versioning} rather than through externalized decoupling.

\noindent\textbf{AWS KMS.} Operations use uniform function signatures, but algorithm identity leaks through required parameters: the \texttt{Sign} operation requires a \texttt{SigningAlgorithm} argument that the caller must specify. This places operations at C1.1. Key creation uses named key specifications (\texttt{KeySpec=ECC\_NIST\_P256}), which function as named algorithm identifiers analogous to template names (C2.1).

AWS KMS supports multiple key store types with identical API surface (C3.2, C4 at 1 on C3). IAM policies provides role-based access authority (C5.2): distinct policies can grant distinct permissions. Such policies can constrain the templates that may be used, either for key creation or cryptographic operations, with a \textit{kms:KeySpec} condition. It achieves limited cryptographic governance: one can specify \textit{who} may perform \textit{which operations} with \textit{which algorithm} but cannot specify \textit{which algorithms are allowed} nor enforce compliance with regulations. These are key requirements for achieving complete cryptographic governance and precisely the difference between C5.2 and C5.3+. Enablers are limited to the same-algorithm key rotations (E1.1) for algorithm migration and manual import (E2.1) for provider migration.

{\emergencystretch=1em
\noindent\textbf{HashiCorp Vault Transit.} The \texttt{/transit/sign/:name} and \texttt{/transit/encrypt/:name} endpoints dispatch entirely on the named key's configuration, with no algorithm parameters in the request (C1.2). Key creation uses named type identifiers (C2.1, C4 at 0 on C2).\par}

Enterprise Vault supports managed keys backed by HSMs or cloud KMS services, with identical API surface and automatic routing (C3.3). Vault ACL policies provide role-based access control (C5.2), but govern access rather than cryptographic selection. Key rotation is only for the same algorithm (E1.1), and keys can be exported and imported (E2.1).

\subsection{Synthesis and Gap Analysis}

Three pervasive and independent patterns emerge from this evaluation. Each of these patterns identifies a gap that is sufficient, independently, to prevent an agile migration.

\noindent\textbf{Pattern 1: Operation decoupling is partial, and creation decoupling is absent.} All the six APIs we evaluated achieve at least C1.1 (uniform function signatures). However, four of the six (PKCS\#11, OpenSSL, JCA, and AWS~KMS) remain at level C1.1. This means algorithm-specific parameters still leak through operation interfaces. Only Tink achieves C1.4 (fully managed) and Vault achieves C1.2 (key-driven dispatch).

Importantly, \textit{no evaluated system achieves C2.2 (intent-based) key creation}. All six APIs require algorithm-specific identifiers when creating keys. None of them  support expressing cryptographic intent during key creation. This means that adopting a new algorithm requires modifying every key-creation call site. This gap alone makes the post-quantum transition a per-site code-change problem across all evaluated systems.

\noindent\textbf{Pattern 2: Access authority exists, but cryptographic governance does not.} AWS KMS and Vault both reach C5.2 (role-based access authority) level for authority support. However, no evaluated system achieves C5.3+ (authority over cryptographic algorithm selection) for cryptographic governance and control.

\noindent\textbf{Pattern 3: Enablers are the missing dimension.} No system reaches E1.3 (explicit algorithm transformation). Only Tink achieves E1.2 (multi-algorithm versioning). AWS KMS and Vault reach E1.1 (same-algorithm rotation), and PKCS\#11, OpenSSL, and JCA provide no versioning (E1.0) support. Provider migration is uniformly weak across all the evaluated API (E2 at 0 or 1).

This gap has a concrete consequence: even when systems are reasonably decoupled, they cannot leverage that decoupling for algorithm migration. An organization using Vault Transit with ECDSA P-256 keys cannot transform those keys to ML-DSA-65. It must create entirely new keys, update all references, re-sign all data, and decommission old keys.

\noindent\textbf{The Orthogonal Advantage.} The component-based framework reveals agility profiles that a single linear score would obscure. Tink achieves C1.4 and E1.2 but only C2.1, C3.0, and C4.0. Vault achieves C1.2 and C3.3 but only C2.1 and C4.0 on C2. Ranking one system as ``more agile'' than another requires specifying the dimension of the ranking.

PKCS\#11 and JCA share identical C1 (1) and C2 (0) but differ on C3 (2 vs.\ 1) and E2 (1 vs.\ 0). AWS KMS and Vault share identical C2 (1) and C5 (2) but differ on C1 (1 vs.\ 2) and C3 (2 vs.\ 3). These non-hierarchical profiles demonstrate that cryptographic agility comprises genuinely independent dimensions.

\subsection{Threats to Validity}

Several factors may affect the validity of our framework and evaluation.

\textit{Construct validity.} The framework's seven dimensions were derived from analysis of deployed systems rather than from formal specification. Alternative decompositions are possible, and we chose the current decomposition because it captures the coupling patterns most consequential for the post-quantum transition and produces discernible assessments.

\textit{Internal validity.} Assessments relied on public APIs, official documentation, and available source code. Undocumented capabilities may shift the individual scores we measure. However, the cross-cutting patterns, such as the universal absence of C2.2+, C5.3+, and E1.3, are robust to minor score adjustments.

\textit{External validity.} We chose six systems spanning the three API generations and diverse architectural paradigms (hardware abstraction, cryptographic libraries, language frameworks, opinionated toolkits, and managed services) for evaluation. Broader coverage may reveal additional patterns. The three patterns identified are likely to generalize because they reflect structural properties of existing API paradigms.

\textit{Reliability.} Component-level assignments involve judgment calls at boundary cases, and we mitigated this by defining precise level criteria and documenting our reasoning for each assignment.

\section{Derived Design Requirements}
\label{sec:requirements}

The evaluation reveals that the three gaps, absent intent-based creation, absent cryptographic governance, and absent algorithm transformation, are not independent implementation omissions but consequences of missing architectural provisions. This section derives the design requirements from the framework's analysis to close these gaps. These requirements are developed into concrete API design principles and Protocol Buffers patterns in the companion paper~\cite{paper2-api}.

\subsection{Requirements from the Coupling Dimensions}

\noindent\textbf{R1: Algorithm-independent operations.} Achieving C1.2+ requires that cryptographic operations carry no algorithm identity in their invocation. At C1.2, the system resolves the algorithm from the key's metadata, removing algorithm names and system-generatable parameters from the call site. At C1.3+, user-supplied operational parameters must additionally be scope-consistent across all algorithms within a scope (intent). Operation signatures must be defined by cryptographic goals (encrypt, sign, derive) and must accept only key identifiers, user data, and scope-level parameters.

\noindent\textbf{R2: Intent-based key creation vocabulary.} Achieving C2.2+ requires a scope vocabulary through which applications express cryptographic intent without naming algorithms. This vocabulary must guarantee \textit{operational parameter consistency}: all algorithms within a scope must accept identical parameters from the application during a cryptographic operation. This will ensure that algorithm substitution does not cause runtime failures. Scope boundaries must therefore be drawn where user-supplied operational parameters diverge.

\noindent\textbf{R3: Stable key identifiers.} Supporting C1.2 (key-driven dispatch), E1.1+ (versioning), and E2.1+ (migration) simultaneously requires key identifiers for representing keys. The key identifier must not encode algorithm identity, and must retain its identity across algorithm rotations, provider migrations, and policy updates.

\noindent\textbf{R4: Provider-algorithm separation.} Achieving C3.2+ requires a structural separation between algorithm specifications and provider implementations. Algorithm properties (such as security strength and mathematical structure) are determined by the algorithm's specification and do not vary across implementations. Provider properties (such as FIPS~140-3 certification level, side-channel resistance, and hardware isolation) are deployment-specific facts about particular implementations. The same algorithm (e.g., AES-256-GCM) may be implemented by a software library without FIPS certification, by a FIPS~140-3 Level~3 certified HSM, or by a formally verified library. Conflating algorithm and provider properties couples algorithm selection to provider selection, preventing independent evolution of either concern.

\subsection{Requirements from the Decoupling Mechanism}

\noindent\textbf{R5: Temporal decoupling.} Achieving C4.1+ requires that decisions made at different times (code authorship, deployment, runtime) be changeable independently. It must be possible to change algorithms after deployment without modifying code-authorship-time decisions in the source code.

\noindent\textbf{R6: Pluggable policy engine.} Achieving C4.3+ requires a policy engine whose implementation is not prescribed by the API. The API should support governance-related aspects without mandating a specific schema for policy specification.

\subsection{Requirements from the Authority Dimension}

\noindent\textbf{R7: Separation of concerns.} Achieving C5.3+ requires that the API support distinct roles: developers who invoke operations, security architects who define approved algorithms, and operations teams who configure providers. Each role must be able to perform its function independently.

\subsection{Requirements from the Agility Enablers}

\noindent\textbf{R8: Three-stage key evolution.} Achieving E1.1--E1.3 requires three distinct evolution operations: \textit{rotation} (new key material generated under the same algorithm), \textit{transformation} (new key material generated under a different algorithm), and \textit{provider migration} (existing key material moved to a different execution backend). All three must preserve the logical key identifier and maintain a complete version history to support decryption or verification of data protected under prior versions.

\noindent\textbf{R9: Discoverable extensibility.} Supporting the addition of new algorithms (including post-quantum) without API changes requires a template registry with lifecycle management and a discovery API that allows runtime querying of available algorithms and their properties.
These nine requirements collectively address all seven framework dimensions and identify the minimal architectural provisions needed to close the three gaps. The companion paper~\cite{paper2-api} develops these requirements into design principles organized under foundational architectural properties. The paper also demonstrates it through concrete Protocol Buffers API patterns.

\section{Related Work}
\label{sec:related}

\subsection{Defining Cryptographic Agility}

N\"ather~et~al.~\cite{towards_common_understanding} confirm that multiple interpretations of crypto-agility exist in the literature, including crypto-agility as a system property, engineering practice or design paradigms. The authors propose a canonical definition, specifying crypto-agility as capabilities for identifying and modifying cryptographic methods in a flexible and efficient way while preserving business continuity. NIST \cite{NIST_considerations_agility} differentiates crypto-agility depending on its context (computing systems, protocol, or organization) to clearly define the meaning of business continuity and requires to integrate crypto agility into organizations' governance with, for example, cryptographic policies to control the algorithms and libraries allowed for use, or enforce compliance with standards and regulations. RFC~7696~\cite{rfc7696} characterizes cryptographic agility for protocols as the ability to migrate from one mandatory-to-implement algorithm suite to another over time. Our work builds on these foundations to decompose agility into independently assessable dimensions for API-level agility requirements.

\subsection{Assessment Frameworks}

{\emergencystretch=1em
The Crypto-Agility Maturity Model~\cite{maturity-model} defines four levels of agility readiness that include requirements for cryptographic providers, such as extensibility, decoupling of application and cryptography, and policies to restrict allowed algorithms. Authors of \cite{agility-finance} extends the maturity model to the financial sector, providing a topology for crypto agile architecture suggesting defining cryptography in a service layer. The maturity models emphasize agility for IT systems, but lacks concrete requirements for cryptographic APIs. Our framework complements it by specifying an assessment method for cryptographic providers.\par}

CARAF~\cite{caraf} quantifies the risk and cost associated with the lack of cryptographic agility. This risk-oriented framework complements our capability-oriented framework: CARAF measures the \textit{consequences} of insufficient agility, while our framework measures the \textit{architectural provisions} that determine agility capability.

Our component-based model fills the gap between organizational-level frameworks and concrete API design by providing fine-grained, independently assessable dimensions with precise level definitions.

\subsection{Application-Level Agility in APIs}

Tink~\cite{tink2018} introduces primitive-level interfaces that achieve the highest operation decoupling (C1.4) among deployed systems, but does not provide intent-based key creation, externalized algorithm selection, or provider portability. SecAlgo~\cite{secalgo} allows selecting the cryptographic library at configuration time, providing provider abstraction that Tink lacks. EverCrypt~\cite{evercrypt} enables automated multiplexing across formally verified implementations depending on the target hardware to optimize for performance. Its primitive-level interface lacks a common interface for digital signatures, prescribing operational uniformity for digital signature schemes. eUCRITE~\cite{eucrite} addresses algorithm abstraction through predefined security levels (LOW, MEDIUM, HIGH) bound to default algorithm in source code, which requires modifying and redeploying the library to change default algorithms. 

Among these, none provides intent-based key creation (C2.2+), policy-driven algorithm selection (C4.3+), or explicit algorithm transformation (E1.3+). Our framework identifies these gaps as structural limitations of current API design paradigms.

\subsection{Systems-Level Agility}

ELCA~\cite{elca} introduces a centralized cryptographic configuration service conceptually similar to our C4.3 applied to a microservice architecture. Cho~et~al.~\cite{software-defined-crypto} propose a software-defined cryptography layer applying SDN-inspired principles to cryptographic provisioning. Both demonstrate the viability of the architectural separation our framework advocates.

Our work complements these systems-level approaches by defining the agility requirements that the underlying cryptographic provider APIs must satisfy. A complete agility stack requires both: provider APIs that support intent-based creation and key evolution (addressed in the companion paper~\cite{paper2-api}), combined with orchestration layers for policy distribution and migration coordination.

\subsection{Standards and Guidelines}

NIST SP~800-57~\cite{nist-sp800-57} provides key lifecycle guidance but assumes fixed algorithm choices at key creation---an assumption our framework challenges by introducing C2.2+ and E1.3+. FIPS~203~\cite{fips203}, 204~\cite{fips204}, and 205~\cite{fips205} specify post-quantum algorithms but not migration strategies. NIST SP~1800-38~\cite{nist-sp1800-38} provides practical migration guidance, and IETF PQUIP~\cite{ietf-pquip} addresses protocol-level adoption, but neither defines a systematic assessment framework for application-level agility.

\subsection{Security and Usability of APIs}

Decades of API evolution have produced operation-level abstraction and provider models, but intent-based abstraction and policy-driven configuration remain unrealized. Green and Smith~\cite{green-smith-usability} argue that APIs should make the secure choice the easy choice; our framework extends this principle by arguing that the \textit{agile} choice---the design decision that preserves future algorithm substitutability---should be systematically assessable. Lazar~et~al.~\cite{lazar2014} and Acar~et~al.~\cite{acar2017} document that API design significantly influences security outcomes; our framework addresses agility as a complementary architectural concern sharing the same root cause.

\subsection{Policy Languages and Governance}

XACML~\cite{xacml} and OPA~\cite{opa} address access control rather than cryptographic selection. Our framework's distinction between access authority (C5.2) and cryptographic governance (C5.3+) makes explicit that authorization frameworks, regardless of their sophistication, cannot provide cryptographic governance without an intent vocabulary (C2.2+) and an externalized selection mechanism (C4.3+).

CycloneDX~\cite{cyclonedx} version~1.6+ defines a Cryptography Bill of Materials (CBOM). Our framework suggests that systems achieving C2.2+ possess the metadata necessary for automated CBOM generation.

\section{Conclusion}
\label{sec:conclusion}

\subsection{Summary of Contributions}

This paper introduced a component-based assessment framework for cryptographic agility that decomposes a frequently conflated concept into seven orthogonal dimensions: three coupling dimensions (C1 operation coupling, C2 creation coupling, C3 provider coupling), a cross-cutting decoupling mechanism (C4), a governance authority dimension (C5), and two agility enablers (E1 algorithm migration, E2 provider migration). The framework captures non-hierarchical agility profiles, enabling precise characterization of what kind of agility a system provides and what kind it lacks.

We evaluate six representative cryptographic APIs against the framework and identify three pervasive gaps for cryptographic agility:

\begin{enumerate}
\item \textbf{Partial operation decoupling, absent creation decoupling.} Most systems achieve uniform operation interfaces (C1.1+), but all six require algorithm-specific identifiers at key creation (C2.0 or C2.1). No system supports expressing cryptographic intent without explicitly specifying an algorithm (C2.2+).

\item \textbf{Access authority without cryptographic governance.} Two systems (AWS~KMS, Vault) provide role-based access control (C5.2), but no system achieves authority over cryptographic algorithm selection (C5.3+). The framework's C5 dimension makes the conflation between access control and cryptographic governance precise and assessable.

\item \textbf{Absent algorithm transformation.} No system provides explicit algorithm transformation (E1.3), and only one (Tink) achieves multi-algorithm versioning (E1.2). Consequently, even where decoupling exists, it cannot be exercised for cross-algorithm migration.
\end{enumerate}

These gaps are independently sufficient to prevent agile migration and mutually reinforce the agility gap. Together, they explain why the post-quantum transition is a software development problem despite decades of progress in cryptographic API design.

\end{document}